\documentclass[letterpaper]{mn2e}
\usepackage{epsfig,natbib2,natbibmnfix}

\newcommand{\be}{\begin{equation}}
\newcommand{\ee}{\end{equation}}

\newcommand{\apj}{ApJ}
\newcommand{\apjs}{ApJS}
\newcommand{\mnras}{MNRAS}
\newcommand{\aap}{A\&A}

\newcommand{\apjl}{ApJL}

\newcommand{\nat}{Nature}

\def\ltsima{$\; \buildrel < \over \sim \;$}
\def\simlt{\lower.5ex\hbox{\ltsima}}
\def\gtsima{$\; \buildrel > \over \sim \;$}
\def\simgt{\lower.5ex\hbox{\gtsima}}

\newcommand\fe{Fe K$\alpha$\ }

\def\del#1{{}}

\title{Anisotropic X-ray emission in AGN accretion discs.}

\author[S.~Nayakshin]
{\parbox{18cm}{Sergei Nayakshin\footnotemark[1]}\vspace{0.3cm}\\
Department of Physics \& Astronomy, University of Leicester, Leicester, LE1 7RH, UK}

\begin{document}

\maketitle

\begin{abstract}
Straight-forward models of X-ray reflection in the inner region of accretion
discs predict that primary X-ray flux and the flux reflected off the surface
of the disc should vary together, albeit a short light travel time delay. Most
of the observations, however, show that the X-ray flux can vary while the
reflected features remain constant. Here we propose a simple explanation to
this. In all likelihood, the emission of a moderately optically thick magnetic
flare atop an accretion disc is anisotropic. A constant energy release rate in
a flare will appear to produce a variable X-ray flux as the flare rotates with
the accretion disc anchoring the magnetic tube. The reflector, on the other
hand, receives a constant X-ray flux from the flare. Since the reflected
emission is azimuthally symmetric, the observer will see a roughly constant
reflected flux (neglecting relativistic effects). The model does not produce
quasi-periodic oscillations (QPO) if magnetic flux tubes are sheared out
faster than they complete one orbit.
\end{abstract}

\begin{keywords}
{Galaxy: centre -- accretion: accretion discs -- galaxies: active}
\end{keywords}
\renewcommand{\thefootnote}{\fnsymbol{footnote}}
\footnotetext[1]{E-mail: {\tt Sergei.Nayakshin at astro.le.ac.uk}}

\section{Introduction}
\label{intro}

Rapid X-ray variability of accreting black holes \citep[e.g., see the case for
AGN in][]{Done89} suggests that in many of these sources X-rays are emitted
very close to the last stable orbit of an accretion disc presumed to power the
emission. At the same time, most of the power emitted by AGN
\citep[e.g.,][]{Elvis94} seems to come out in the optical/UV band, implying
that the disc is relatively cold all the way to the last stable orbit. X-rays
are thus emitted in a physical proximity to the cold disc. X-rays impinging on
any cold matter should produce fluorescent emission lines and the continuum
"reflection component'' from AGN \citep{Lightman88,Guilbert88}. The latter
feature has indeed been found in observations almost two decades ago
\citep[e.g.,][]{Pounds90}. Due to a high fluorescent yield and abundance of Fe
among heavy elements in the interstellar medium, the strongest fluorescent
line is expected to be the \fe line \citep[e.g.,][]{George91}.

The reflected spectral components are also expected to be relativistically
shifted and broadened, which should be especially noticeable for the case of
the \fe line \citep{FabianEtal89}.  Indeed, broad \fe lines are rather common
in the time-averaged observed spectra of AGN \citep[e.g.,][]{Fabian00}. At the
same time, there is a considerable number of AGN without broad \fe lines. In
the view of the constraints on the physical location of X-ray and optical/UV
emitting materials mentioned above, we interpret this fact as evidence for
extreme ionisation of the accretion disc surface in the bright
"reflection-free" sources\footnote{Low Luminosity AGN are known to lack the
thermal optical/UV bump in their spectral energy distributions \citep{Ho99}.
These sources are likely to lack broad \fe lines since cool accretion discs do
not appear to extend all the way to the last stable orbit
\citep{PtakEtal04}.}.  In particular, \cite{NKK, NayakshinKallman01} have
shown that if the X-rays are emitted in localised, very bright magnetic
flares, then the disc regions next to them are over-ionised. These regions are
covered by optically thick "skin" of temperature $T \sim$ few keV, where Fe
and other spectroscopically important metals are completely ionised. The skin
thus does not produce the atomic features expected in the less ionised X-ray
reflection models. It is our view that time-averaged X-ray observations of
AGN, with and without broad \fe lines, can all be explained in terms of
photo-ionised X-ray reflection models \citep{Nayakshin00b}. Unfortunately,
spectral ambiguity of time-averaged models is usually large as the parameter
space of the models is large \citep{BarrioEtal03}. Based on such spectra alone
one could also suggest that in narrow \fe line AGN the innermost region of the
disc is truncated and is replaced by a hot quasi-spherical corona, as might be
the case in hard state of stellar-mass black hole binary systems
\citep{Esin97}.

Time-resolved \fe line and X-ray reflection observations are a powerful tool
to learn more detail about the geometry in immediate proximity of accreting
black holes than possible from time-averaged spectra. If X-rays are indeed
emitted by separate bright but small and localised magnetic flares, one
expects narrow but relativistically shifted \fe lines in snapshot spectra of
AGN \citep{NayakshinKazanas01}. Such lines have been observed in a few cases
\citep[e.g.,][]{IwasawaEtal04,TurnerEtal04,TurnerEtal06} despite the
challenges of the limited photon statistics, broadly confirming the basic
picture of an X-ray source above an cold disc extending very close to the
black hole.

However, time-resolved X-ray spectroscopy brought a number of unexpected
surprises that present serious challenges to the X-ray reflection/reprocessing
paradigm. A robust theoretical prediction of the model is that a change in the
X-ray continuum luminosity should lead to a corresponding variation in the
luminosity of the reflected features albeit with a small delay time. This
model prediction has not been confirmed in most of the observations suitable
for such an analysis.

A lack of fast variability in the \fe line was reported by
\cite{RevnivtsevEtal99} for a famous X-ray binary containing a black hole
Cyg-X1. \cite{Vaughan01} showed that \fe line in Seyfert 1 galaxy MCG 6-30-15
does vary on the expected short time scales, but this variability is not
correlated to the changes in the X-ray continuum flux.
\cite{MarkowitzEtal03}, using observations of 7 AGN with the {\em RXTE}, find
that ``there is no evidence for correlated variability between the line and
continuum, severely challenging models in which the line tracks continuum
variations modified only by a light-travel time delay''. They find a decreased
amplitude in the \fe line variability as compared to that in the X-ray
continuum, similar to the \cite{RevnivtsevEtal99} results for Cyg-X1.

Observations of other manifestations of X-ray reprocessing pose similar
problems. \cite{MiniuttiEtal06}, using a recent {\em Suzaku} observations of
the Seyfert 1 galaxy MCG 6-30-15, find that variability in the 14-45 keV
energy band is quenched with respect to that at a few keV.  They further find
that this is robustly explained by a two-component model, in which the
reflection bump, dominating the harder energy band, remains constant while a
power-law component varies rapidly. \cite{ReevesEtal06} finds similar results
for two other sources, MCG-5-23-16 and NGC 4051. \cite{EdelsonEtal00} finds a
similarly puzzling absence of a short term variability in the optical that
would be correlated to the rapidly varying X-ray emission in the Seyfert
galaxy NGC 3516. These authors conclude that their observations challenge the
standard X-ray reprocessing paradigm where X-ray emission is partially
reflected but mainly reprocessed \citep{Lightman88,Guilbert88} into the
optical/UV bands.

Time-dependent changes in the structure of the ionised disc atmosphere may
decouple the X-ray flux and the reflected spectrum
\citep[e.g.,][]{NK02,CollinEtal03,CzernyEtal04}, but a situation in which the
reflected continuum appears constant would be very fine tuned.
\cite{MiniuttiEtal06} propose that such nominally unexpected spectral
behaviour can be explained by the relativistic light bending model
\citep{Miniutti04}, in which the X-ray emitting source is located above the
black hole. In this model the intrinsic luminosity of the source is constant
but its height above the black hole is changing with time. Due to strong
relativistic effects, the observed flux can vary by more than one order of
magnitude, whereas the reflected spectrum is far less variable as the
relativistic effects are not as severe. \cite{MalzacEtal06} confirm the
relativistic bending results of \cite{Miniutti04}, and propose another model,
in which the reflecting medium is highly inhomogeneous.

In this paper we explore an alternative and rather simple explanation for the
worrying uncorrelated variability of the X-ray continuum and the reflected
features. We argue that X-ray emission from realistic magnetic flare
structures should be expected to be anisotropic. These flares must rotate with
the disc material as the magnetic flux tubes are anchored into the disc, and
hence a distant observer would see the flare under a time-dependent
angle. Unless the emitting region is (very) optically thin, the observer then
witnesses a varying X-ray flux even if the angle-integrated power output in
the flare is constant. At the same time, the reflected spectrum would remain
constant because this component should be axially symmetric.

\section{The argument}\label{sec:model}

Magnetic flares occurring on the surface of the Sun are believed to have very
complex geometry \citep[e.g.,][]{DemoulinEtal97,NishioEtal97,LiEtal00}. In
general, emission region consists of several flux tubes possibly interacting
with each other. There is no spherical or axial symmetry in this case. It is
hard to see why magnetic flares on the surface of an accretion disc would be
any more symmetric than Solar flares.

We shall now argue that even if emission within the source is locally
isotropic, the resulting emission is anisotropic for a mildly optically thick
flare. Consider a line of sight passing through an emitting region to the observer.
The standard radiative transfer equation shows that the specific radiation
intensity emerging from the region is
\begin{equation}
I = I_0 \exp\left[-\tau\right] + \int_0^{\tau} d\tau' S(\tau') \exp\left[-\tau'\right]\;,
\label{general}
\end{equation}
where $\tau$ is the total optical depth of the region along the line of sight,
$I_0$ is the radiation intensity entering the region, and $S(\tau')$ is the
source function. We assume that the emitting region is located above the cold
disc that emits no X-rays \citep{Haardt93}, and hence we shall set $I_0=0$ in
the X-ray domain.

For an optically thin source, $\tau\ll 1$, we have
\begin{equation}
I \approx \int_0^{\tau} d\tau' S(\tau') = \int_{\rm inside} \frac{d
x'}{\lambda} S(\tau')\;,
\label{tau0}
\end{equation}
where $\lambda$ is the photon mean free path, $dx' \equiv \lambda d\tau'$, and
the integral is taken over $x'$ as long as $x'$ is within the source. The
X-ray flux from the source is obtained by integrating equations 1 or 2 over
the whole projected surface area of the source visible to the
observer. Equation 2 yields
\begin{equation}
F =  \int_{\rm inside} d^3 {\bf r}\, \frac{S(\bf{r})}{\lambda} \;,
\label{full}
\end{equation}
where ${\bf r}$ is the 3D coordinate and the integration is now taken over the
three-dimensional volume of the source (we omit the solid angle of the
detector as seen from the source; this factor is obviously constant).  If the
source function is isotropic, we see that the flux emitted in any direction is
the same for an optically thin source. This conclusion holds for an arbitrary
geometry of the source as long as $\tau \ll 1$.

However, radiation transfer models require Thomson optical depths of the order
of $\tau \sim 1$ to explain the observed X-ray continuum
\citep{Haardt93}. In this case, the approximate equality \ref{tau0} is no
longer valid. The amount of radiation emitted in a given direction depends on
the typical value of the optical depth $\tau$ in that direction, even if the
source function itself is isotropic,

Detailed polarised radiation transfer calculations for cylindric and
hemispheric emitting regions were performed by \cite{Poutanen96}. The axis of
symmetry of the regions were perpendicular to the reflector (disc) in these
calculations. The results showed, among other important conclusions, that the
resulting X-ray emission strongly depends on the viewing angle. For example,
their Figure 2, panel b, shows that emission of a hemisphere with Thomson
optical depth $\tau$ as small as 0.07 is very different for a pole-on and an
edge-on views. It is very obvious to us that if the axis of symmetry of the
hemisphere were tilted away from the symmetry axis of the disc, the emission
would also depend on the azimuthal angle $\phi$ defined in the plane of the
reflector.

We conclude that if magnetic flares on the surface of the disc are anything
like those observed on the surface of the Sun and have $\tau\simlt 1$, then
X-ray emission flux is guaranteed to be anisotropic. More specifically, the
emission will depend on both the inclination angle of the disc to the observer
and the orbital phase of the flare as seen by the observer.  At the same time,
the emission from the atmosphere of the disc illuminated by X-rays from above
\citep[e.g.,][]{Ross93,ZyckiEtal94,NKK} is not expected to depend on the
azimuthal angle (neglecting relativistic effects). The reflected emission does
depend on the inclination angle of the disc \citep[e.g.][]{NKK}, but this
angle is not expected to vary for a given source.

Hence, if a constant energy release rate magnetic flare rotates together with
the disc region underneath it, the observer will see a varying X-ray continuum
flux and a constant X-ray reflection flux.

\section{Conclusions}

We made a very simple point in this paper, emphasising the potential
importance of anisotropy of X-ray emitters in accreting black holes.  One
prediction of this model is an uncorrelated X-ray variability of the continuum
and the reflected features on time scales shorter than a dynamical time
($R^{3/2}/(GM)^{1/2}$, where R is the radius at the location of the flare, and
$M$ is the black hole mass), as observed in a number of cases.

We do not expect the model to be applicable to variability on time scales
longer than the dynamical time. Magnetic loop's foot-points are expected to
follow the motion of the differentially rotating disc. One therefore suspects
the loops to be significantly deformed on the local dynamical time. Therefore,
magnetic flux tube's emissivity is likely to evolve on this time scale rather
than remain constant. In other words, on time scales longer than a local
dynamical time, variability in the total number of flares or their emissivity
would be more important than the effects we discussed.

Our model for the uncorrelated X-ray variability of the continuum and the
reflected features could also work if an anisotropic X-ray source is placed on
the disc symmetry axis, as in the light bending model of \cite{Miniutti04}. In
that case the source must rotate (i.e. due to the black hole spin) but can
remain at a constant height above the black hole.

The author acknowledges illuminating discussions with Simon Vaughan.

\bibliographystyle{mnras}

\end{document}